\documentclass[aps,prl,twocolumn,showpacs,amsmath,amssymb]{revtex4}

\usepackage{graphicx}
\usepackage{dcolumn}
\usepackage{bm}

\begin{document}
\title{
Anomalous Hall Effect and Skyrmion Number
in Real- and Momentum-space
}
\author{Masaru Onoda$^1$}
\author{Gen Tatara$^2$}
\author{Naoto Nagaosa$^{1,3}$}
\affiliation{
$^1$Correlated Electron Research Center (CERC),
National Institute of Advanced Industrial Science and Technology (AIST),
Tsukuba Central 4, Tsukuba 305-8562, Japan\\
$^2$Graduate School of Science, Osaka University, Toyonaka, Osaka 560-0043,
Japan\\
$^3$Department of Applied Physics, University of Tokyo,
Bunkyo-ku, Tokyo 113-8656, Japan}
\date{\today}

\begin{abstract}
We study the anomalous Hall effect (AHE) for the double exchange
model with the exchange coupling $|J_H|$ being smaller than the
bandwidth $|t|$ for the purpose of clarifying the following
unresolved and confusing issues:
(i) the effect of the underlying lattice structure,
(ii) the relation between AHE and the
skyrmion number, (iii) the duality between real and momentum spaces,
and  (iv) the role of the disorder scatterings; which is more essential,
$\sigma_H$ (Hall conductivity) or $\rho_H$ (Hall resistivity)?
Starting from a generic expression for $\sigma_H$, we resolve
all these issues and classify the regimes in the parameter space of
$J_H \tau$ ($\tau$: elastic-scattering time), and
$\lambda_{s}$ (length scale of spin texture).
There are two distinct mechanisms of AHE; one is
characterized by the real-space skyrmion-number,
and the other by momentum-space skyrmion-density at the Fermi level, which
work in different regimes of the parameter space.
\end{abstract}

\maketitle

The anomalous Hall effect (AHE) is a phenomenon where
the Hall resistivity has an additional contribution 
due to the spontaneous magnetization in ferromagnets.  
This anomalous contribution has been attributed to the 
spin-orbit interaction, and various mechanisms has been proposed
\cite{PR095_001154_54,Phy024_00039_58,PTP27_000772_62,JP034_000901_73}.
Recently it has been recognized that the original expression
by Karplus and Luttinger \cite{PR095_001154_54}, i.e., 
the intrinsic contribution,  has 
the geometrical meaning in terms of the Berry-phase curvature 
in momentum space \cite{JPSJ71_00019_02,PRL88_207208_02,Sci302_00092_03}.
This is analogous to the 
the integer quantum Hall effect (IQHE) with the strong
external magnetic field \cite{PRL49_000405_82,AP160_000343_85}.
It was also proposed that AHE arises even without the spin-orbit 
interaction if the spin configuration is non-coplanar
with finite spin chirality, i.e.,
the solid angle subtended by the spins where the electron hops successively
\cite{PRL83_003737_99,PRL84_000757_00,PRB62_R06065_00,Sci291_02573_01,PRL91_057202_03}.
Consider the double-exchange model
\begin{equation}
H = \sum_{\langle \bm{r},\bm{r}'\rangle}
t\;\bm{c}^{\dagger}_{\bm{r}}
\bm{c}_{\bm{r}'}
-\frac{J_H}{2}\sum_{\bm{r}}\bm{S}_{\bm{r}} \cdot
\left[\bm{c}^{\dagger}_{\bm{r}}
\bm{\sigma}\bm{c}_{\bm{r}}\right]
\label{eq:H}
\end{equation}
where $\langle \bm{r},\bm{r}'\rangle$ runs the nearest neighbor sites,
$\bm{c}^{(\dagger)}_{\bm{r}}= 
(c^{(\dagger)}_{\bm{r}\uparrow}, c^{(\dagger)}_{\bm{r}\downarrow})$
is the annihilation (creation) operator at the site $\bm{r}$,
and $\bm{S}_{\bm{r}}$ is the classical spin localized at the site $\bm{r}$.
Assuming a strong Hund coupling $|J_H|(\gg|t|)$  
between the conduction electrons and the localized spins, 
the Berry phase of the localized spins acts as a fictitious magnetic field 
for the conduction electron
\cite{PR100_000675_55,PRB39_011413_89,PRB46_005621_92}.
Ye \textit{et al.} assumed that this fictitious magnetic field 
has a uniform component due to the spin-orbit interaction 
in the presence of the uniform magnetization \cite{PRL83_003737_99}.
However there is a subtle issue concerning the definition of 
the skyrmion number when the spins are defined on the discrete points 
and/or the underlying lattice is relevant.
This is related to the length scale with respect to 
the spin texture and/or the lattice structure.
Furthermore, the effect of the spin-orbit interaction can not be represented 
by the spatially uniform magnetic field; it induces the effective 
``magnetic field'', i.e., the Berry phase curvature, in \textit{momentum} space.
In real systems, the disorder is also relevant and often the following
question arises: Which is more essential, the Hall conductivity $\sigma_H$
or the Hall resistivity $\rho_H$?
Therefore it is highly desirable to resolve all these issues 
in a unified fashion by clearly articulating
the connection between the AHE and the skyrmion number.
In this paper, we study the AHE for the double exchange model eq.~(\ref{eq:H})
in the case of the small exchange coupling $|J_H|$ compared with 
the bandwidth $|t|$.

It is found that the non-coplanar spin-configuration
induces the AHE through the two distinct mechanisms originated by 
($\mathrm{M}_\mathrm{I}$) the non-zero topological-windings of spin texture,
and ($\mathrm{M}_\mathrm{II}$) the nontrivial structure of underlying lattice.
Here the nontrivial structure means that the Wigner-Seitz unit cell
contains multiple sites and different kinds of plaquettes.
Although the two mechanisms work simultaneously in generic cases,
we can identify the two distinct limits where these mechanisms
works exclusively:

\noindent
$\mathrm{C}_\mathrm{I}$: The unit cell contains a single site and
the length scale $\lambda_s$ of the spin texture 
is sufficiently longer than the underlying lattice constant $a$ 
($\lambda_s \gg a$).

\noindent
$\mathrm{C}_\mathrm{II}$: The unit cell contains multiple sites and
the periodicity of the spin texture is the same as the underlying lattice
($\lambda_s =a$).

The mechanism $\mathrm{M}_\mathrm{I}$ is dominant 
and the AHE is characterized by the 
real-space skyrmion-number in the case $\mathrm{C}_\mathrm{I}$, 
while $\mathrm{M}_\mathrm{II}$ is dominant and 
is characterized by the momentum-space skyrmion-density 
at the Fermi level in $\mathrm{C}_\mathrm{II}$.
In the latter case, the AHE takes place 
even if the spin texture has no winding in real space
and the total skyrmion-number is zero in momentum space.

We also discuss the effect of disorder which causes
the following crossover.
The Hall conductivity $\sigma_{H}$ is proportional 
to $|J_{H}^{3}|\tau^{2}/|t|$  for 
$|J_{H}| \ll |t| (a/{\lambda_{s}})^{d} \ll 1/\tau 
\ll |t|(a/{\lambda_{s}})$ in the case $\mathrm{C}_\mathrm{I}$,
and for $|J_{H}| \ll 1/\tau$ in the case $\mathrm{C}_\mathrm{II}$.
Here $\tau$ is the elastic-scattering time, 
and $d$ is the dimensionality of a system.
Hence the Hall resistivity, 
$\rho_{H} = -\sigma_{H}/(\sigma_{0}^2+\sigma_{H}^2)$ 
($\sigma_{0} \propto \tau$, $\sigma_{0}\gg |\sigma_{H}|$),
does not depend on $\rho_{0}\cong 1/\sigma_{0}$,
where $\sigma_{0}$ and $\rho_{0}$ are
diagonal conductivity and resistivity respectively.
However, when $1/\tau \ll |J_{H}|$,
$\sigma_{H}$ approaches its intrinsic value, and
$\rho_{H}$ is proportional to $\rho_{0}^2$
even in the case where $|J_{H}|$ is sufficiently smaller than the band width.
This suggests that the intrinsic (i.e., not due to
impurity scattering) meaning of the AHE 
can be observed through $\rho_{H}$ \cite{Science303_1647_04}
even in the weak coupling regime.

In order to consider the present issues
with taking into account the effect of disorder,
it is transparent to start with the St\v{r}eda formula 
\cite{JPC10_002153_77},
\begin{eqnarray}
\sigma_{\mu\nu}
&=&
-\frac{1}{2\pi V}\int dE f_{F}(E)
\nonumber\\
&&\times
\left\langle\mathrm{Tr}
\left[J_{\mu}\frac{dG_{+}}{dE}(E)J_{\nu}G_{\delta}(E)
+\mathrm{H.c.}\right]
\right\rangle_{\mathrm{imp}},
\end{eqnarray}
where $V$ the system volume,
$\bm{J}$ the current operator,
$G_{\pm}(E)=[E-H-H_{\mathrm{imp}}\pm i \eta]^{-1}$,
$G_{\delta}=G_{+}-G_{-}$, 
$H_{\mathrm{imp}}$ being the interaction with impurities,
and $\langle \cdots\rangle_{\mathrm{imp}}$
represents the ensemble average over impurity configuration.
Here and hereafter we take the unit where $c=\hbar=1$.

In this paper, we shall focus on only metallic states ($|t|\tau \gg 1$),
and just replace $G_{\pm}$ by $[E-H\pm i (2\tau)^{-1}]^{-1}$
to bring in the effect of disorder,
i.e. assuming the isotropic impurity potential
and neglecting the localization effect.
At zero temperature and using the symmetry
$\sigma_{\nu\mu}|_{\mu\neq\nu} = -\sigma_{\mu\nu}|_{\mu\neq\nu}$,
the Hall conductivity is given by
\begin{eqnarray}
\sigma_{H}
&\cong&
-\frac{i}{V}\sum_{\alpha,\alpha'}
\bm{e}_{\perp}\cdot
\left[
\langle \bm{\psi}_{\alpha}|\bm{J}
|\bm{\psi}_{\alpha'}\rangle
\times
\langle \bm{\psi}_{\alpha'}|\bm{J}
|\bm{\psi}_{\alpha}\rangle
\right]
\nonumber\\
&&\times\frac{1}{\pi}
\Biggl[
\frac{\arctan(-2\tau\xi_{\alpha})-\arctan(-2\tau\xi_{\alpha'})}
{(\xi_{\alpha}-\xi_{\alpha'})^2}
\nonumber\\
&&+
\frac{2\tau(1+4\tau^{2}\xi_{\alpha}\xi_{\alpha'})}
{(\xi_{\alpha}-\xi_{\alpha'})
[1+(2\tau\xi_{\alpha})^2][1+(2\tau\xi_{\alpha'})^2]}
\Biggr]
\nonumber\\
&\cong&
-\frac{i}{V}\sum_{\alpha,\alpha'}
\bm{e}_{\perp}\cdot
\left[
\langle \bm{\psi}_{\alpha}|\bm{J}
|\bm{\psi}_{\alpha'}\rangle
\times
\langle \bm{\psi}_{\alpha'}|\bm{J}
|\bm{\psi}_{\alpha}\rangle
\right]
\nonumber\\
&&\times
 \frac{\tau^{2}
\left[f^{\tau}_{F}(E_{\alpha})-f^{\tau}_{F}(E_{\alpha'})\right]}
{1+\left[(E_{\alpha}-E_{\alpha'})\tau\right]^{2}},
\label{eq:generic}
\end{eqnarray}
where
$\xi_{\alpha} = E_{\alpha}-\mu_{0}$ ($\mu_0$ : the chemical potential),
$f^{\tau}_{F}(E)$ is $f_{F}(E)$ with replacing
the inverse temperature by $64\tau/(3\pi)$.
The unit vector $\bm{e}_{\perp}$ is normal to the plane
determined by the Hall measurement.
It is noted that $\sigma_H$ 
depends on $\bm{e}_{\perp}$.
The second formula in eq.~(\ref{eq:generic}) 
is an approximation for the first one
by using physical implication, and
it is numerically confirmed that this approximation is
almost exact with the accuracy less than 5\% around peaks.
In the clean limit, $\tau\to\infty$, 
eq.~(\ref{eq:generic}) reduces to the formula obtained in 
Refs.~\cite{PRL49_000405_82,AP160_000343_85}.

Now we shall consider the AHE in the weak-coupling regime
by assuming a periodic configuration of localized spins.
Every lattice-site is classified into sublattices $I=A, B, C\cdots$.
The localized spins satisfy $\bm{S}_{\bm{r}\in {I}} = \bm{S}_I$.
Then the coupling with the localized spins, $H'$, 
is expressed as 
\begin{equation}
H'= -\frac{J_H}{2}
\sum_{I=A,B,C,\cdots}\sum_{\bm{k}}^{\mathrm{1st BZ}}
\bm{S}_{I}\cdot
\left[\bm{c}_{I\bm{k}}^{\dagger}
\bm{\sigma}
\bm{c}_{I\bm{k}}\right],
\label{eq:H'}
\end{equation}
where $\bm{c}_{I\bm{k}}^{(\dagger)}$ is a Fourier transformation
of the annihilation (creation) operator at $I$-sublattice, 
$\bm{c}_{\bm{r}\in{I}}^{(\dagger)}$.
The system has multiple bands even if we start with a single band at $J_{H}=0$.
The band separation is of the order of 
$|t|(a/\lambda_{s})/(k_F\lambda_{s})^{d-1}$ 
except for the spin splitting of the order of $|J_{H}|$,
where $k_F$ is the Fermi momentum in the case $J_H = 0$ and 
$d$ is the dimensionality of the system.
For simplicity, we shall consider the case $k_F\sim 1/a$ hereafter.
When the parameter $J_{H}$ satisfies $|J_{H}|\ll |t|(a/\lambda_{s})^{d}$
and $\bm{k}$ is not near level crossings,
approximated eigenvalues and eigenfunctions are obtained 
by the conventional perturbation theory for spin-degenerate systems,
\begin{eqnarray}
E_{n\bm{k}\pm} &=& 
E_{n\bm{k}} \mp \frac{J_H}{2}|\bm{S}_{n\bm{k}}|
+\cdots,
\label{eq:Enk}
\\
\bm{S}_{n\bm{k}} &=& \sum_{I=A,B,C,\cdots}\bm{S}_{I}|u_{n\bm{k},I}|^2,
\label{eq:eff-spin}
\\
|\bm{\psi}_{n\bm{k}\pm}\rangle
&=& \sum_{n'}\sum_{s=\pm}\sum_{l=0}
c^{(l)}_{[n'\bm{k}s][n\bm{k}\pm]}
 |\bm{\psi}^{(0)}_{n'\bm{k}s}\rangle
+\cdots,
\label{eq:psi}
\\
|\bm{\psi}^{(0)}_{n\bm{k}\pm}\rangle
&=& \sum_{I=A,B,C,\cdots}\sum_{\sigma=\uparrow,\downarrow}
u_{n\bm{k},I}\chi_{n\bm{k}\pm,\sigma}
c^{\dagger}_{I\bm{k}\sigma} |0\rangle,
\label{eq:psi0}
\end{eqnarray}
where
$c^{(l)}_{[n'\bm{k}s'][n\bm{k}s]}$ is the perturbative coefficient
of the $l$-th order,
$\bm{u}_{n\bm{k}} = (u_{n\bm{k},A},u_{n\bm{k},B},u_{n\bm{k},C},\cdots)^{T}$
is the orbital part of an eigenfunction in $J_H=0$, and
$\bm{\chi}_{n\bm{k}\pm}
=(\chi_{n\bm{k}\pm,\uparrow},
\chi_{n\bm{k}\pm,\downarrow})^{T}$
is the spin part of an eigenfunction satisfying
$
[\bm{S}_{n\bm{k}}\cdot\bm{\sigma}]\bm{\chi}_{n\bm{k}\pm}
=\pm |\bm{S}_{n\bm{k}}|\bm{\chi}_{n\bm{k}\pm}.
$
It is noted that $\bm{S}_{n\bm{k}}$ is regarded as
the effective spin which is felt by the $n$-th pair of bands.
Here, we use the word ``the $n$-th pair of bands'' for the bands with indices
$[n\bm{k}+]$ and $[n\bm{k}-]$.
Below we shall consider the case $\mathrm{C}_\mathrm{I}$ and
$\mathrm{C}_\mathrm{II}$ separately based on the above results.\\

\noindent
\underline{\textit{Case} $\mathrm{C}_\mathrm{I}$}

In this case, the orbital part of an eigenfunction is given by
$
\bm{u}_{n\bm{k}} =
 \bm{u}_{n} =
1/\sqrt{N_{\mathrm{sub}}}
(1, e^{i\bm{b}_{n}\cdot\bm{a}_{B}}, e^{i\bm{b}_{n}\cdot\bm{a}_{C}},\cdots)^{T},
$
where $N_{\mathrm{sub}}$ is the number of sublattices,
$\bm{b}_{n}$ is a reciprocal lattice vector,
$\bm{a}_{I}$ is a lattice vector between $I$-sublattice
and $A$-sublattice.
It is noted that $\bm{u}_{n\bm{k}}$ has no $\bm{k}$-dependence,
and both $\bm{S}_{n\bm{k}}$ and $\bm{\chi}_{n\bm{k}\pm}$
do not depend on the index [$n\bm{k}$] in this case.
Using eqs.~(\ref{eq:Enk})-(\ref{eq:psi0}), 
$\sigma_{H}$ is estimated as  
\begin{eqnarray}
\sigma_{H}
&\cong&
-\frac{1}{V}\sum_{\bm{k}}^{\mathrm{1st BZ}}
\sum_{n, n', n''}'
\bm{e}_{\perp}\cdot
\biggl[\bm{\nabla}E_{n\bm{k}}\times\bm{\nabla}E_{n'\bm{k}}
\nonumber\\
&&
+\bm{\nabla}E_{n'\bm{k}}\times\bm{\nabla}E_{n''\bm{k}}
+\bm{\nabla}E_{n''\bm{k}}\times\bm{\nabla}E_{n\bm{k}}\biggr]
\nonumber\\
&&\times
\frac{J_{H}^3
\Re\left[\bm{S}_{nn'}\cdot(\bm{S}_{n'n''}\times\bm{S}_{n''n})\right]}
{4(E_{n\bm{k}}-E_{n'\bm{k}})
(E_{n'\bm{k}}-E_{n''\bm{k}})(E_{n''\bm{k}}-E_{n\bm{k}})}
\nonumber\\
&&\times
\frac{\tau^{2}
\left[f^{\tau}_F(E_{n\bm{k}})-f^{\tau}_F(E_{n'\bm{k}})\right]}
{1+\left[(E_{n\bm{k}}-E_{n'\bm{k}})\tau\right]^{2}},
\label{eq:single-band}
\end{eqnarray}
where 
$
\bm{S}_{nn'}
= \sum_{I}u^{*}_{n\bm{k},I}\bm{S}_{I}u_{n'\bm{k},I}
= \sum_{I}u^{*}_{n,I}\bm{S}_{I}u_{n',I}
$,
and the intra-pair contribution ($n=n'$) has been neglected
because it is of the order of $J_{H}^{5}$.
The symbol $\sum_{n, n',n''}'$ represents 
the summation with the condition that
every index is different from each other.
More precisely, $\bm{S}_{nn'}$ in eq.~(\ref{eq:single-band})
is expressed as $\bm{S}_{[n\bm{k}][n'\bm{k}]}$,
and $\bm{S}_{n \bm{k}}$ in eq.~(\ref{eq:eff-spin})
is $\bm{S}_{[n\bm{k}][n\bm{k}]}$.
However, in the case $\mathrm{C}_\mathrm{I}$,
$\bm{S}_{[n\bm{k}][n'\bm{k}]}$ has no 
$\bm{k}$-dependence.

Changing the relative scales of
the parameters $|t|$, $|J_H|$, $1/\lambda_{s}$ and $1/\tau$,
$\sigma_{H}$ shows the crossover.
Here we shall consider the following two typical cases.

\noindent
$\mathrm{C}_\mathrm{I}$-$\mathrm{A}$:
$|J_{H}|\ll |t|(a/\lambda_{s})^{d} \ll 1/\tau \ll |t|(a/\lambda_{s})$.
This means that the length scale $\lambda_s$ is shorter 
than the elastic-scattering length $\ell\sim |t|a\tau$, 
i.e. $\lambda_{s}\ll \ell$. In this case 
$
|\sigma_{H}| \propto |J_{H}|^{3}\tau^{2}/|t|.
$

\noindent
$\mathrm{C}_\mathrm{I}$-$\mathrm{B}$:
$1/\tau \ll |J_{H}| \ll |t|(a/\lambda_{s})^{d}$,
i.e., the level separation is sufficiently larger than $1/\tau$,
where $\sigma_{H}$ takes its intrinsic value.

In the former case $\mathrm{C}_\mathrm{I}$-$\mathrm{A}$,
especially in two dimensions,
we can derive the topological meaning of the AHE
by relating $\sigma_{H}$ to the skyrmion number.
Because of the condition $|J_{H}|\ll |t|(a/\lambda_{s})^{d} \ll 1/\tau$,
we can approximate the summations of band indices in eq.~(\ref{eq:single-band})
by energy integrals, and estimated them by 
residues $E=\mu_{0}\pm i/(2\tau)$ in the complex energy-plane.
(See the first formula of eq.~(\ref{eq:generic}).)
By the inverse Fourier transformation to real-space variables,
we can obtain the expression equivalent to 
that given by Tatara and Kawamura \cite{JPSJ71_02613_02}
for the periodic spin-configuration.
They identified that $|J_H| \tau$ is the small parameter for
the perturbative expansion of $\sigma_{H}$.
Hence finite $\tau$ is essential there. 
The Hall conductivity was shown to be proportional to a sum of
the spin chirality of any three localized spins with a geometrical weight 
in real space. 
\begin{eqnarray}
\sigma_{H} 
&\cong& \frac{e^2}{2\pi}\frac{J_H^3\tau^2}{|t|}\chi,
\label{eq:TK}
\end{eqnarray}
where $\chi$ is the total chirality in real space,
\begin{eqnarray}
  \chi &=&  \frac{a^{4}}{V}\sum_{\bm{r}_{i}}
     \frac{\bm{e}_{\perp}\cdot(\bm{l}\times\bm{l}')}{ll'} 
I'(l) I'(l') I(l'')
\nonumber\\
&&\times
\bm{S}_{\bm{r}_{1}}\cdot(\bm{S}_{\bm{r}_{2}}\times\bm{S}_{\bm{r}_{3}})
, 
\label{chiunidef}
\end{eqnarray}
$\bm{l}=\bm{r}_{1}-\bm{r}_{2}$,
$\bm{l}'=\bm{r}_{2}-\bm{r}_{3}$ and 
$\bm{l}''= \bm{r}_{3}-\bm{r}_{1}$.
$I(r) \propto \sum_{\bm{k}}e^{i\bm{k}\cdot\bm{r}}
/[1+(2\tau\xi_{\bm{k}})^{2}]$ 
is an RKKY-type function which decay exponentially
in the scale of $\ell$ because of the complex part
of the residues in energy integrals, and
$I'(r)=\frac{d I(r)}{dr}$.
Contribution from largely separated three spins
with the scale of $r$ decays rapidly as
$\sim e^{-3r/2\ell}$, and the AHE is dominantly driven 
by chiralities of spins on small triangles.
Using the expansion
$\bm{S}_{\bm{r}_1}\cong \bm{S}_{\bm{r}_2}
+\bm{l}\cdot\bm{\nabla}\bm{S}_{\bm{r}_2}$,
$\bm{S}_{\bm{r}_3}\cong \bm{S}_{\bm{r}_2}
-\bm{l}'\cdot\bm{\nabla}\bm{S}_{\bm{r}_2}$,
we obtain
\begin{equation}
\chi= \frac{S^3 A}{N}
\int d^2\bm{r} \Phi^z (\bm{r}),
\end{equation}
where $S = |\bm{S}_{\bm{r}}|$,
$N$ is the number of lattice sites,
\begin{equation}
\Phi^z(\bm{r}) = \frac{1}{4\pi S^3}
\bm{S}\cdot(\nabla_x\bm{S} \times\nabla_y \bm{S})
\end{equation}
is the skyrmion density in real space, and
\begin{equation}
A \sim \frac{1}{a^4}
\int^{\lambda_{s}}  dl \int^{\lambda_{s}} dl'
l^{2}l'^{2}
I'(l) I'(l') I(l'').
\end{equation}
Because of $\lambda_{s} \ll \ell$,
$A$ is a dimensionless function of $\lambda_{s}/a$,
when higher order terms in $\lambda_{s}/\ell$ is neglected.
Thus the Hall conductivity is given by
\begin{equation}
\sigma_{H} \cong \frac{e^2}{2\pi}\cdot\frac{J_{H}^{3}\tau^{2}}{|t|}
\cdot\frac{S^3 A N_{s}}{N},
\label{eq:real-space}
\end{equation}
where $N_{s}=\int d^{2}\bm{r}\Phi^z(\bm{r})$ is the skyrmion number
\cite{ICM, Zhou}.\\

\noindent
\underline{\textit{Case} $\mathrm{C}_\mathrm{II}$}

Here the unit cell contains multiple sites and
the periodicity of spin texture is the same as the underlying lattice.
In this case, the band separation is of the order of $|t|$
except for the spin degeneracy and the degeneracy at the symmetric points.
The orbital part of the eigenfunction $\bm{u}_{n\bm{k}}$
has non-trivial $\bm{k}$-dependence,
and the same goes for $S_{n\bm{k}}$ and $\bm{\chi}_{n\bm{k}}$.
As we shall see below, 
the intra-pair contribution, which contains matrix element
$\langle \bm{\psi}_{n\bm{k}\mp}|\bm{J}|
\bm{\psi}_{n\bm{k}\pm}\rangle$, 
is of the order of $|J_H|^3\tau^2/|t|$ or $|J_H|/|t|$
depending on $|J_{H}|\tau \ll 1$ or $\gg 1$.
The inter-pair contribution, which contains the matrix element
$\langle \bm{\psi}_{n\bm{k}s}|\bm{J}|
\bm{\psi}_{n'\bm{k}s'}\rangle$ ($n\neq n'$), 
is of the order of $|J_H|^3/|t|^3$, and thus
negligibly small compared with the intra-pair contribution.
Therefore, the dominant contributions to $\sigma_{H}$
come from the intra-pair terms in sharp contrast to 
the case $\mathrm{C}_\mathrm{I}$. 

Considering that the $n$-th pair of bands 
is intersecting the Fermi level,
the dominant contribution is given by
\begin{eqnarray}
\sigma_{H}
&\cong& 
-\frac{2i}{V}\sum_{\bm{k}}^{\mathrm{1st BZ}}
\bm{e}_{\perp}\cdot\left[
\langle\bm{\psi}_{n\bm{k}+}|
\bm{J}
|\bm{\psi}_{n\bm{k}-}\rangle
\times
\langle\bm{\psi}_{n\bm{k}-}|
\bm{J}
|\bm{\psi}_{n\bm{k}+}\rangle
\right]
\nonumber\\
&&\times
\frac{\tau^{2}\left[f^{\tau}_F(E_{n\bm{k}+})-f^{\tau}_F(E_{n\bm{k}-})\right]}
{1+\left[(E_{n\bm{k}-}-E_{n\bm{k}+})\tau\right]^{2}}.
\end{eqnarray}
The perturbative eigenfunctions give the following 
matrix element of the current operator
\begin{equation}
\langle\bm{\psi}_{n\bm{k}\mp}|
\bm{J}
|\bm{\psi}_{n\bm{k}\pm}\rangle
\cong
-\frac{e}{2}J_H|\bm{S}_{n\bm{k}}| 
\bm{\chi}^{\dagger}_{n\bm{k}\mp}
\left[
\bm{\nabla}_{\bm{k}}\bm{e}_{n\bm{k}}\cdot\bm{\sigma}
\right]
\bm{\chi}_{n\bm{k}\pm},
\label{eq:current-matrix}
\end{equation}
where the higher order terms in $J_H/t$ are neglected, and
$\bm{e}_{n\bm{k}}=\bm{S}_{n\bm{k}}/|\bm{S}_{n\bm{k}}|$.

Using the above result and the completeness condition,
\begin{equation}
\sum_{s=+,-}
\bm{\chi}_{n\bm{k}s}\otimes
\bm{\chi}^{\dagger}_{n\bm{k}s}
=
\left(
\begin{array}{cc}
1 & 0 \\
0 & 1
\end{array}
\right),
\end{equation}
we can estimate $\sigma_{H}$ as follows,
\begin{eqnarray}
\sigma_{H}
&\cong& \frac{e^2}{2V}\sum_{\bm{k}}^{\mathrm{1st BZ}}
\left[-\frac{df^{\tau}_F}{dE}(E_{n\bm{k}})\right]
\bm{e}_{\perp}\cdot\bm{\Phi}_{n\bm{k}}
\nonumber\\
&&\times
\frac{(J_H|\bm{S}_{n\bm{k}}|)^{3}\tau^{2}}
{1+\left(J_H|\bm{S}_{n\bm{k}}|\tau\right)^{2}},
\label{eq:weak}
\end{eqnarray}
where
\begin{equation}
\Phi^{\lambda}_{n\bm{k}}
=
\frac12
\sum_{\mu,\nu}
\epsilon^{\lambda\mu\nu}
\bm{e}_{n\bm{k}}\cdot
\left(
\nabla_{k_{\mu}}\bm{e}_{n\bm{k}}
\times
\nabla_{k_{\nu}}\bm{e}_{n\bm{k}}
\right).
\label{eq:chirality}
\end{equation}
It is noted that $\bm{\Phi}_{n\bm{k}}$ may be regarded as
the effective chirality of the $n$-th pair of bands in momentum space,
because its integration is related to the solid angle of $\bm{S}_{n\bm{k}}$.
Especially in a two-dimensional system, 
$\bm{e}_{\perp}\cdot\bm{\Phi}_{n\bm{k}}/(4\pi)$
represent the skyrmion density in momentum space.
This means that
$\sigma_{H}$ is characterized by the momentum-space
skyrmion-density at the Fermi level.
When we change the parameter $|J_{H}|\tau$,
$\sigma_{H}$ shows the crossover,
\begin{eqnarray}
|\sigma_{H}| \propto \left\{
\begin{array}{ll}
\frac{|J_{H}|^{3}\tau^{2}}{|t|}, &
\left(|J_{H}|\tau\ll 1\right) \\
\frac{|J_{H}|}{|t|}, &
\left(|J_{H}|\tau\gg 1\right)
\end{array}
\right. .
\label{eq:crossover}
\end{eqnarray}
Finally, it is noted that, in contrast to the case $\mathrm{C}_\mathrm{I}$,
the AHE takes place even if there is no winding of spin texture in real space
and the momentum-space skyrmion-number is also zero.

\begin{figure}[h]
\includegraphics[scale=0.4]{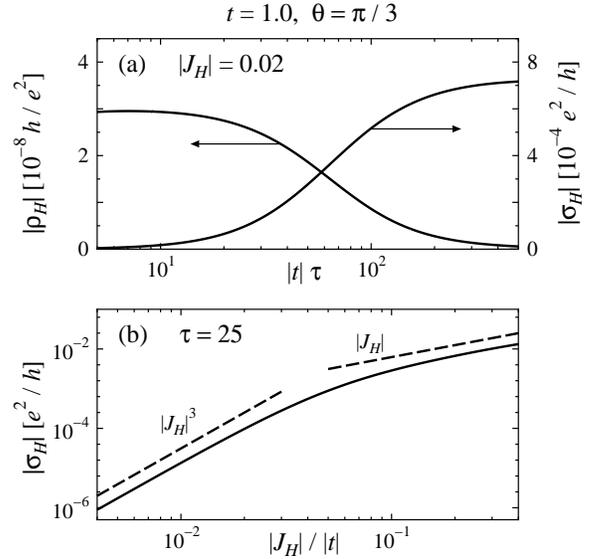}
  \caption{
(a) $\tau$-dependence of
$\rho_{H}$ and $\sigma_{H}$,
and (b) $|J_{H}|$-dependence of $\sigma_{H}$
for the kagome-lattice model in the case where
the average number of electrons per unit cell is three.
It is shown in the panel (a) that
$\rho_{H}$ and $\sigma_{H}$ approach to constant values 
for $|J_{H}|\tau \ll 1$ and for $|J_{H}|\tau \gg 1$
respectively.
The panel (b) shows
the crossover from $|\sigma_{H}|\propto|J_{H}|^{3}$ for $|J_{H}|\tau \ll 1$
to $|\sigma_{H}|\propto|J_{H}|$ for $|J_{H}|\tau \gg 1$.
}
\label{fig:sigma-rho}
\end{figure}

In order to confirm the above consideration 
for the case $\mathrm{C}_\mathrm{II}$,
we present the explicit results in Fig.~\ref{fig:sigma-rho}
for the model eq.~(\ref{eq:H}) 
on the kagome lattice with the spin texture,
\begin{equation}
\frac{\bm{S}_I}{S} = \left(\sin\theta\cos\phi_{I},
 \sin\theta\sin\phi_{I},
\cos\theta\right),
\label{eq:kagomeS}
\end{equation}
where $\phi_{I} = (4n_{I}-1)\pi/6$ 
($n_{A} = 1, n_{B} = 2$, $n_{C} = 3$).
The results are calculated 
by using the first line of eq.~(\ref{eq:generic}),
and clearly shows the crossover as predicted in eq.~(\ref{eq:crossover}).

In conclusion, we have shown that 
the non-coplanar spin configuration induces the AHE
by two distinct mechanisms.
The AHE is characterized by the real-space 
skyrmion-number when the underlying lattice structure is irrelevant
and the spin texture is slowly varying as $a\ll \lambda_{s}\ll \ell$.
On the other hand, the AHE is
characterized by the momentum-space skyrmion-density at the Fermi level
when the underlying lattice structure is relevant and 
the periodicity of the spin texture is the same as the lattice.
The Hall resistivity and conductivity
become essential, i.e., independent of the elastic-scattering time,
for $|J_{H}|\tau \ll 1$ and for $|J_{H}|\tau \gg 1$ respectively.

\begin{acknowledgements}
The authors would like to thank K.~Ohgushi and M.~Yamanaka
for useful discussion.
M.~O. is supported by Domestic Research Fellowship
from Japan Society for the Promotion of Science.
G.~T. thanks the Mitsubishi foundation for financial support.
N.~N. is supported by NAGERI and Priority Areas Grants and Grant-in-Aid 
for COE research  
from the Ministry of Education, Science, Culture and  Sports of Japan.
\end{acknowledgements}



\end{document}